\documentclass[journal=jctcce,manuscript=article]{achemso}

\usepackage[T1]{fontenc}
\usepackage{siunitx, chemformula}
\usepackage{multirow}
\usepackage{enumitem}
\usepackage[colorlinks=true, citecolor=blue, urlcolor=blue, linkcolor=blue, breaklinks=true, pdfpagelabels=false]{hyperref}

\newcommand*\m[1]{\mathrm{#1}}

\newcommand*\msi[2]{\SI[group-digits=false]{#1}{#2}}
\newcommand*\mb[1]{\mathbf{#1}}
\newcommand*\dd[1]{\mathrm{d}{#1}}

\newcommand*\ket[1]{|{#1}\rangle}

\newcommand*\expval[1]{\langle{#1}\rangle}
\newcommand*\annotation[1]{\textsuperscript{\emph{#1}}}

\author{Xiaoyu Zhang}
\altaffiliation{Contributed equally to this work}
\author{Tai Wang}
\altaffiliation{Contributed equally to this work}
\email{wtpeter@pku.edu.cn}
\author{Yi Qin Gao}
\email{gaoyq@pku.edu.cn}
\author{Yunlong Xiao}
\affiliation{New Cornerstone Science Laboratory, College of Chemistry and Molecular Engineering, Peking University, Beijing 100871, China}
\email{xiaoyl@pku.edu.cn}

\title{Noncollinear Spin-Flip TDDFT for Potential Energy Surface Crossings: Conical Intersections and Spin Crossings}

\begin{document}

\begin{tocentry}
\includegraphics[width=3.25in]{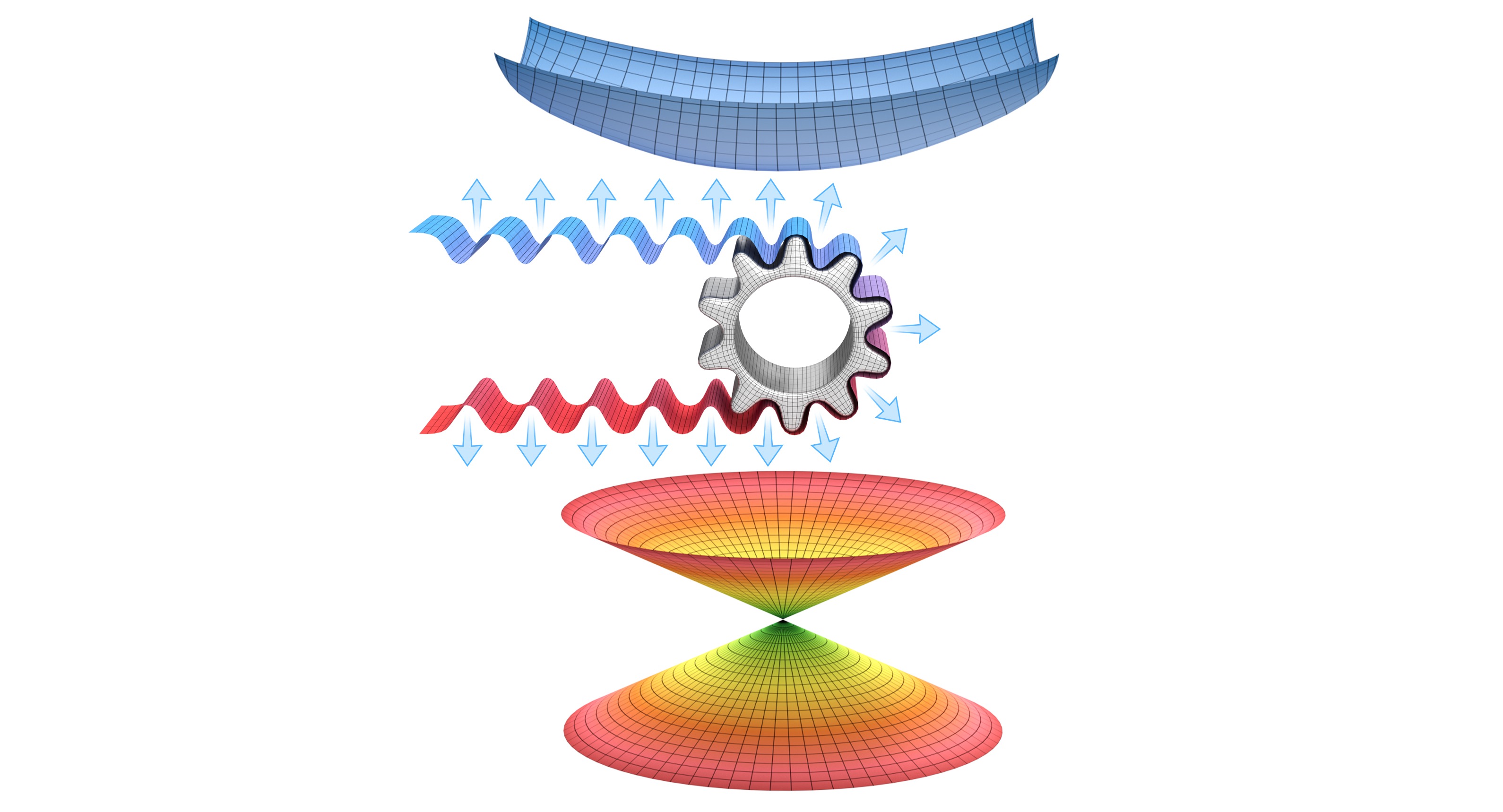}
\end{tocentry}

\begin{abstract}
We recently proposed a scheme to generalize collinear functionals to the noncollinear regime, termed the multicollinear approach, which preserves spin symmetry and provides numerically stable higher-order functional derivatives. This scheme has been applied to noncollinear spin-flip TDDFT and its analytic gradient calculations. In the present work, with the aid of the penalty function method, we employ the noncollinear spin-flip TDDFT in multicollinear scheme to locate potential energy surface crossings. We investigate two distinct types of crossings and analyze their topographical and spin characteristics near the crossing points. The first type is conical intersections between states of the same spin multiplicity, such as the singlet ground and excited states. The second type involves spin crossings that occur between electronic states with different spin multiplicities, such as between singlet and triplet. These crossing regions enable ultrafast nonadiabatic transitions through either nonadiabatic coupling or spin-orbit coupling, playing a crucial role in photochemistry. Through theoretical analysis and illustrative examples, we demonstrate the advantages of noncollinear spin-flip TDDFT over  conventional collinear spin-flip TDDFT or spin-conserving TDDFT. Finally, we systematically evaluate its prospects as an electronic structure method for use in nonadiabatic molecular dynamics.
\end{abstract}

\maketitle

\section{\label{sec:introduction}Introduction}

The degeneracy of electronic potential energy surfaces (PESs) is pivotal in determining the dynamics of molecular systems, especially in the context of photochemical reactions and nonradiative transitions\cite{schuurman2018dynamics}. At the most fundamental level, considering only the intersections between two states, these crossings can be classified into two types according to their spin multiplicities: conical intersections and spin crossings. Conical intersections occur between states of the same spin multiplicity, such as between two singlet states, and serve as photochemical funnels facilitating ultrafast internal conversion\cite{yarkony1996diabolical,yarkony1999s1s0,domcke2012role,haas2014conical,matsika2021electronic}. In contrast, spin crossings involve states of different spin multiplicities, as exemplified by the intersection between singlet and triplet states, thereby governing intersystem crossing processes via spin-orbit coupling effects\cite{harvey1998singlet,casavecchia2015reaction,vanuzzo2016reaction,li2019intersystem,wang2021conical}.

The dimensionality of the crossing region is a fundamental aspect in understanding these degeneracies\cite{gozem2014shape,ryabinkin2017geometric,matsika2021electronic}. This topic traces its origins to the celebrated non-crossing rules of von Neumann and Wigner\cite{newmann1929behaviour}, and has been extensively studied in various contexts. As discussed by Mead\cite{mead1979noncrossing}, the dimensionality of the intersection depends on whether spin-orbit coupling is present and on whether the system has an even or odd number of electrons. For the simplest case of a non-relativistic, even-electron system, the electronic Hamiltonian representing two diabatic states as a function of nuclear coordinates $\mb{R}$ is given by:
\begin{equation}
    H(\mb{R}) = \begin{pmatrix}
        H_{11}(\mb{R}) & H_{12}(\mb{R}) \\
        H_{21}(\mb{R}) & H_{22}(\mb{R})
    \end{pmatrix},
\end{equation}
where $H_{12}=H_{21}$ and the $2\times 2$ matrix $H$ is real.
The conditions for degeneracy are given by
\begin{gather}
    H_{11}(\mb{R}) = H_{22}(\mb{R}),\\
    H_{12}(\mb{R}) = 0, \label{H12}
\end{gather}
which define a subspace where two PESs are degenerate. For a molecule with $N$ atoms, the number of internal nuclear degrees of freedom is $M = 3N - 6$ (or $3N - 5$ for linear molecules). To simultaneously satisfy the two independent conditions above, two degrees of freedom are constrained, and thus the intersection forms an $(M-2)$-dimensional seam, known as a conical intersection. The remaining two dimensions, referred to as the branching space, lift the degeneracy. In contrast, spin crossings are described by an $(M-1)$-dimensional intersection, as degeneracy is lifted in only one coordinate direction, because Eq.~\ref{H12} is always satisfied for states of different spin multiplicities.

Accurate descriptions of conical intersections often require expensive multireference methods, such as multiconfigurational self-consistent field (MCSCF)\cite{helgaker2000} or multireference configuration interaction (MRCI)\cite{shavitt1977}. In contrast, computationally efficient methods like density functional theory (DFT)\cite{hohenberg1964inhomogeneous,kohn1965selfconsistent} and spin-conserving time-dependent DFT (TDDFT)\cite{runge1984densityfunctional,casida1995timedependent} fail to produce the correct dimensionality of the intersection between the ground state (using DFT) and excited state (using TDDFT), as Eq.~\ref{H12} is automatically satisfied due to Brillouin's theorem\cite{levine2006conical}. To recover the correct topological features of conical intersections between ground and excited states within DFT and spin-conserving TDDFT, additional modifications are required, such as configuration interaction corrections (CIC)\cite{li2014configuration,xu2025conical}, dual functional (DF) methods\cite{shu2017dualfunctional}, particle-particle (pp) or hole-hole (hh) treatments\cite{yang2013double,vanaggelen2013exchangecorrelation,yang2016conical,bannwarth2020hole}, or the explicit inclusion of double excitations (TDDFT-1D)\cite{teh2019simplest}.

\begin{figure}[htb]
    \centering
    \includegraphics[width=0.95\textwidth]{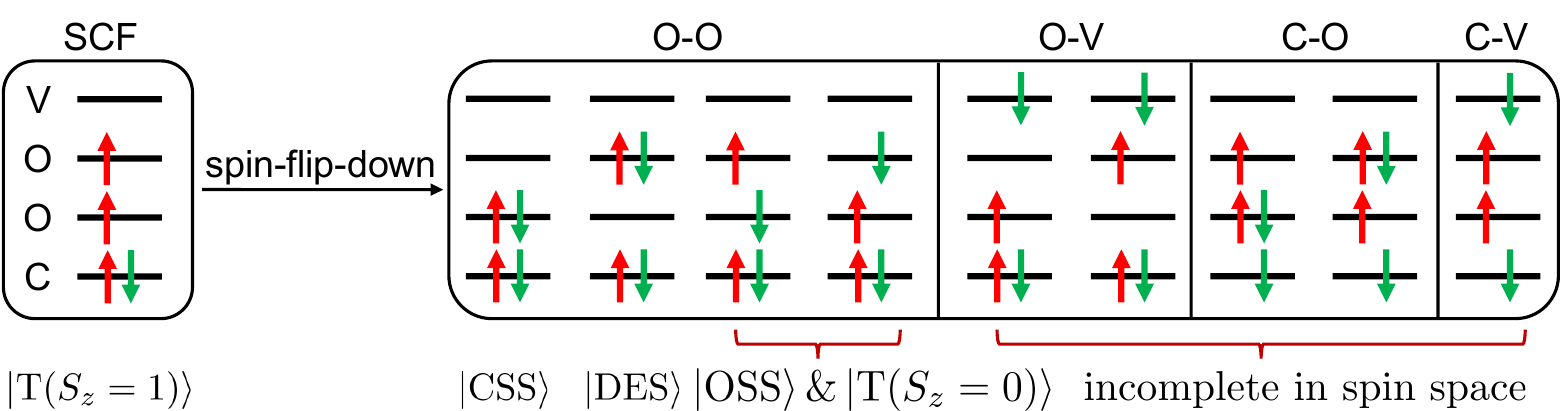}
    \caption{\label{fig:sfsch}Schematic diagram of the spin-flip-down TDDFT starting from a high-spin triplet reference state $\ket{\m{T}(S_z=1)}$. C, O, and V stand for doubly-occupied orbitals, singly-occupied orbitals, and virtual orbitals, respectively. For the O-O type excitations, four spin-complete configurations can be generated, including closed-shell singlet $\ket{\m{CSS}}$, doubly-excited singlet $\ket{\m{DES}}$, open-shell singlet $\ket{\m{OSS}}$, and spin-flip triplet $\ket{\m{T}(S_z=0)}$. The O-V, C-O, and C-V excitations yield states that lack compensating configurations to maintain the spin multiplicity.}
\end{figure}

One promising strategy to restore the correct dimensionality of conical intersections is the adoption of spin-flip methodologies\cite{minezawa2009optimizing,harabuchi2013automated,gozem2014shape,zhang2014analytic,winslow2020comparison}. Depending on the type of excitation, spin-flip TDDFT can be categorized into spin-flip-up and spin-flip-down channels. In this work, we focus on the latter, as shown in Figure~\ref{fig:sfsch}. By introducing single spin-flip-down excitations from a high-spin reference state $\ket{\m{T}(S_z=1)}$, the spin-flip-down TDDFT approach enables a unified description of both the ground and excited states.

When spin-flip TDDFT was first proposed by Krylov \emph{et al.}\cite{shao2003spin}, it employed collinear exchange-correlation functionals and was therefore called collinear spin-flip TDDFT. In such framework, the coupling between molecular orbitals of different spins arises solely from the Hartree-Fock (HF) exchange component in hybrid functionals. Collinear spin-flip TDDFT has since been widely applied to diverse systems, such as diradicals\cite{kishi2007finitefield,hossain2017photoelectron}, triradicals\cite{cristian2004bonding,krylov2005triradicals}, photochemical simulations\cite{minezawa2009optimizing,harabuchi2014dynamics,yue2018performance,zhang2021nonadiabatic}, and single-molecule magnets\cite{valero2011magnetic}. Despite its practical utility, collinear spin-flip TDDFT has several limitations. First, to ensure proper coupling between spin channels, it generally requires hybrid functionals with a large fraction of HF exchange, which restricts the applicability of pure functionals. Second, it suffers from the so-called self-splitting problem, where energies of the reference state $\ket{\m{T}(S_z=1)}$ and the spin-flip state $\ket{\m{T}(S_z=0)}$ differ artificially, introducing ambiguity and inconsistency in determining the triplet energies. Third, it is prone to spin contamination from three sources: the incomplete spin treatment inherent in the collinear spin-flip kernel, the incompleteness of spin space (Figure~\ref{fig:sfsch}), and the spin contamination already present in the high-spin reference state.

On the other hand, noncollinear spin-flip TDDFT has a clearer and more rigorous theoretical foundation. Wang and Ziegler\cite{wang2004timedependent} realized that applying the nonrelativistic version of two-component noncollinear TDDFT\cite{gao2005timedependent} to a collinear system naturally leads to two decoupled equations: one is the conventional spin-conserving TDDFT equation (which only requires a collinear functional), and the other is the noncollinear spin-flip TDDFT equation (which requires a noncollinear functional). Ref.~\citenum{li2025analytic} provides a detailed derivation of this decoupling. A key implication of this framework is that pure functionals can directly couple the two spin channels\cite{bernard2012general}, making the HF exchange an optional, rather than mandatory, component. The spin-flip kernel originally proposed by Wang and Ziegler was based on a noncollinear local density approximation (LDA) functional and has been applied to the description of photochemical reactions\cite{huix-rotllant2010assessment}. However, when extended to functionals beyond LDA such as generalized gradient approximation (GGA) and meta-GGA, it encounters numerical instabilities due to divisions by zero, which limits its broad applicability\cite{bernard2012general,li2012theoretical}. In our recent work, a general and robust framework for extending noncollinear functionals has been developed, called the multicollinear approach\cite{pu2023noncollinear}. This method is applicable to all types of exchange-correlation functionals and provides numerically stable functional derivatives, thereby revitalizing noncollinear spin-flip TDDFT\cite{li2023noncollinear}. Within the multicollinear framework, the calculation of analytic gradients for noncollinear spin-flip TDDFT also becomes simple and natural\cite{li2025analytic}. The computation of derivative couplings is currently underway.

In this work, noncollinear spin-flip TDDFT within the multicollinear approach is used to locate and characterize conical intersections and spin crossings. Furthermore, we evaluate its potential as an electronic structure computational tool for nonadiabatic dynamics simulations.

\section{\label{sec:method}Methodology}

\subsection{\label{sec:sftd}Spin-Flip TDDFT Formalism}

In this section, we present the working equations for noncollinear spin-flip TDDFT. Detailed derivations can be found in Refs.~\citenum{li2025analytic,wang2025zero}.

In the spin-flip-down formalism, the excitation energies $\Omega$ and the corresponding transition amplitudes $X$ are obtained by solving the Casida equation:
\begin{equation}
    \begin{pmatrix}
        H_{\bar{a}i,\bar{b}j} & H_{\bar{a}i,\bar{j}b} \\
        H_{\bar{i}a,\bar{b}j} & H_{\bar{i}a,\bar{j}b}
    \end{pmatrix}\begin{pmatrix}
        X_{\bar{b}j} \\ X_{\bar{j}b}
    \end{pmatrix}=\Omega\begin{pmatrix}
        1&0\\0&-1
    \end{pmatrix}\begin{pmatrix}
        X_{\bar{a}i} \\ X_{\bar{i}a}
    \end{pmatrix}. \label{SFD}
\end{equation}
Here, $a,b$ denote virtual spin-up orbitals, $i,j$ denote occupied spin-up orbitals, and $\bar{a},\bar{b},\bar{i},\bar{j}$ denote the spin-down orbitals. The $H$ matrix is given by
\begin{equation}
    \begin{pmatrix}
        H_{\bar{a}i,\bar{b}j} & H_{\bar{a}i,\bar{j}b} \\
        H_{\bar{i}a,\bar{b}j} & H_{\bar{i}a,\bar{j}b}
    \end{pmatrix} = \begin{pmatrix}
        \delta_{ij}F_{\bar{a}\bar{b}} - \delta_{\bar{a}\bar{b}}F_{ji}&0\\
        0&\delta_{\bar{i}\bar{j}} F_{ba} - \delta_{ab}F_{\bar{i}\bar{j}}
    \end{pmatrix} + \begin{pmatrix}
        K_{\bar{a}i,\bar{b}j} & K_{\bar{a}i,\bar{j}b} \\
        K_{\bar{i}a,\bar{b}j} & K_{\bar{i}a,\bar{j}b}
    \end{pmatrix},
\end{equation}
where $F$ is the Fock matrix, and $K$ is the spin-flip kernel. For simplicity, we will express the $K$ matrix in terms of one-component basis functionals $\{\phi_{\mu}\}$, with the superscripts $\uparrow$ and $\downarrow$ indicating the spin channels. The spin-flip kernel is expressed as
\begin{equation}
    K_{\mu\nu,\rho\sigma}^{\downarrow\uparrow,\downarrow\uparrow} = -C_{\m{HFx}}(\mu\rho|\sigma\nu) + 2\iint \phi_{\mu}^*(\mb{r})\phi_{\nu}(\mb{r}) \frac{\delta^2 E_{\m{xc}}}{\delta m_x(\mb{r})\delta m_x(\mb{r}')} \phi_{\sigma}^*(\mb{r}')\phi_{\rho}(\mb{r}')\dd{\mb{r}} \dd{\mb{r}'}, \label{sfkernel}
\end{equation}
where $C_{\m{HFx}}$ is the fraction of HF exchange, and $\frac{\delta^2 E_{\m{xc}}}{\delta m_x(\mb{r})\delta m_x(\mb{r}')}$ is the functional derivative with respect to the transverse spin magnetization $m_x$.

The second term in Eq.~\ref{sfkernel} originates from the contribution of pure functionals. It naturally appears in noncollinear spin-flip TDDFT but vanishes in collinear spin-flip TDDFT. This term is critically important because it allows the spin-flip kernel to remain nonzero even when pure functionals are used. Consequently, the Casida matrix in the left-hand side of Eq.~\ref{SFD} no longer reduces to a diagonal form, enabling the calculation of excitation energies beyond mere orbital energy differences. Within the multicollinear approach, it can be computed as follows\cite{li2023noncollinear}
\begin{equation}
    \frac{\delta^2 E_{\m{xc}}}{\delta m_x(\mb{r})\delta m_x(\mb{r}')} = \int_0^1 \frac{\delta^2 E_{\m{xc}}^{\m{col}}[n,\tilde{m}_z]}{\delta \tilde{m}_z(\mb{r})\delta \tilde{m}_z(\mb{r}')}\bigg|_{\tilde{m}_z=tm_z} \dd{t}, \label{Exx}
\end{equation}
where the collinear second derivatives are averaged with respect to a scaled magnetization $\tilde{m}_z = t m_z$. As proved by our recent work\cite{wang2025zero}, this approach guarantees the correct degeneracy of triplet states and is readily applicable to any collinear functional as long as its second derivatives are available.

\subsection{\label{sec:ciloc}Locating the Crossing Points}

Minimum energy conical intersections (MECIs) between singlet states, as well as minimum energy crossing points (MECPs) for singlet-triplet crossings, correspond to local minima within the seam where two electronic states are degenerate. In this work, we adopt the penalty function scheme originally proposed by Levine \textit{et al.},\cite{levine2008optimizing} in which the following objective function is minimized:
\begin{equation}
    F_{IJ}(\mb{R};\sigma,\alpha)=\frac{E_I(\mb{R})+E_J(\mb{R})}{2}+\sigma \frac{\Delta E_{IJ}(\mb{R})^2}{\Delta E_{IJ}(\mb{R})+\alpha},
\end{equation}
where $I$ and $J$ label the two adiabatic states of interest, $\Delta E_{IJ} = E_I - E_J$ is their energy gap, $\sigma$ is a penalty prefactor, and $\alpha$ is a smoothing parameter (set to 0.02 a.u. in this work). Optimization of $F_{IJ}$ requires only the energies and energy gradients of the two relevant states, without the need for derivative couplings.

Following the protocol in Ref.\citenum{levine2008optimizing}, we begin with $\sigma = 3.5$ and minimize $F_{IJ}$. If the energy difference $\Delta E_{IJ}$ at the resulting minimum exceeds a preset threshold (0.01 eV in this work), $\sigma$ is increased and the minimization is repeated iteratively until the degeneracy condition is satisfied. The optimization follows the convergence criteria for both step size and gradient norm of the objective function, as stated in Ref.\citenum{levine2008optimizing}.

\section{\label{sec:ci}Noncollinear Spin-Flip TDDFT for PES Crossings}

In this section, we illustrate the capabilities of noncollinear spin-flip TDDFT for investigating PES crossings through several representative applications. First, we locate $\m{S}_0/\m{S}_1$ conical intersections, which is arguably the most important class of crossings. Noncollinear spin-flip TDDFT successfully identifies these intersections with the correct topological structure and significantly reduces the spin contamination observed in collinear spin-flip TDDFT. Second, we show that collinear spin-flip TDDFT can produce artificial PES crossings when pure functionals are used, whereas the noncollinear approach avoids such spurious features. Third and fourth, we consider $\m{S}_0/\m{T}$ and $\m{S}_1/\m{T}$ crossings. With the zero excitation energy theorem\cite{wang2025zero}, noncollinear spin-flip TDDFT provides consistent triplet energies, enabling unambiguous identification of singlet-triplet intersections. Finally, we demonstrate its versatility by describing $\m{D}_0/\m{D}_1$ conical intersection in an odd-electron system. All calculations are performed using the PySCF package\cite{sun2020recent,pyscfforge}.

\subsection{\label{sec:realci}$\m{S}_0/\m{S}_1$ Conical Intersections}

Collinear spin-flip TDDFT has often been employed to locate $\m{S}_0/\m{S}_1$ conical intersections, and previous studies have shown that it generally provides reasonable MECI geometries\cite{minezawa2009optimizing,harabuchi2013automated,gozem2014shape,zhang2014analytic,winslow2020comparison}. However, the resulting $\m{S}_0$ and $\m{S}_1$ states can exhibit significant spin contamination. This contamination introduces ambiguities in the spin multiplicity of the intersection and can adversely affect subsequent calculations, such as derivative couplings.

\begin{figure}[htb]
    \centering
    \includegraphics[width=0.95\textwidth]{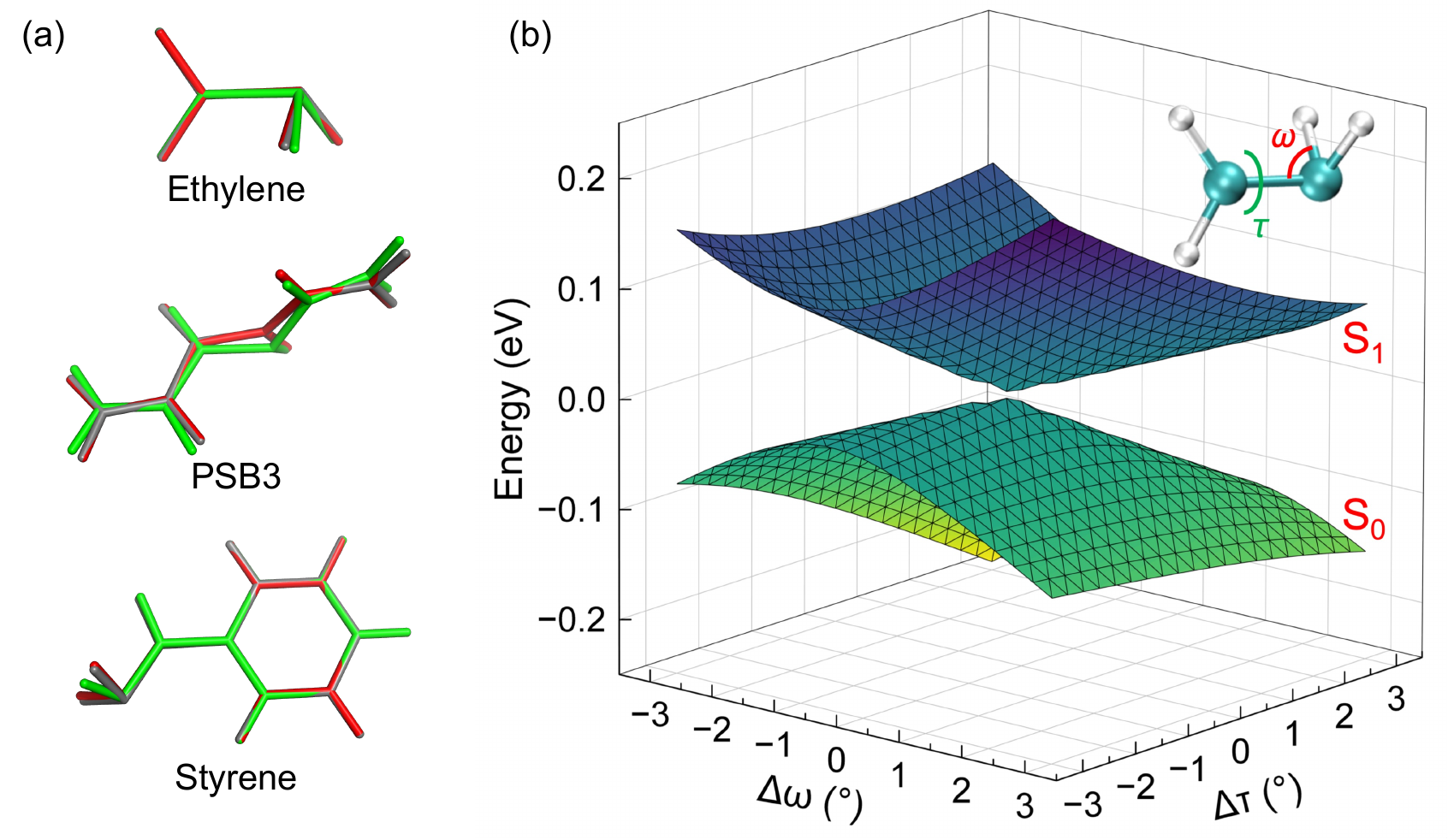}
    \caption{\label{fig:realci}(a) Geometries of MECIs obtained by MRCISD (gray, from Ref.~\citenum{nikiforov2014assessment}), collinear spin-flip TDDFT with BHHLYP (green, from Ref.~\citenum{nikiforov2014assessment}), and noncollinear spin-flip TDDFT with BHHLYP (red, this work). Structures are visualized using PyMOL\cite{PyMOL}. (b) PESs of $\m{S}_0$ and $\m{S}_1$ near the MECI of ethylene. The twisting angles follow the convention in Ref.~\citenum{minezawa2009optimizing}. Structures are visualized using VMD\cite{humphrey1996vmd,stone1998tachyon}.}
\end{figure}

Here, we revisit the benchmark systems of Ref.~\citenum{nikiforov2014assessment}. Three representative molecules--ethylene, penta-2,4-dieniminium cation (PSB3), and styrene--are selected, all of which were previously shown to suffer from significant spin contamination when treated with collinear spin-flip TDDFT. Here, MECIs are optimized for these systems using the BHHLYP functional\cite{becke1988densityfunctional,lee1988development,becke1993densityfunctional} with the 6-31+G** basis set\cite{ditchfield1971selfconsistent,hehre1972self,hariharan1973influence,clark1983efficient}, following the same level of theory as in Ref.~\citenum{nikiforov2014assessment} to ensure direct comparability. The resulting MECI geometries from noncollinear spin-flip TDDFT, collinear spin-flip TDDFT, and MRCISD calculations are illustrated in Figure~\ref{fig:realci}a, with the detailed results summarized in Table~\ref{tab:realci}.

With the same BHHLYP functional, noncollinear spin-flip TDDFT yields structures that are in closer agreement with high-level MRCISD results than those from the collinear approach. It is worth noting that the $\expval{\hat{S}^2}$ values are substantially reduced by approximately half an order of magnitude for both $\m{S}_0$ and $\m{S}_1$ states. These results highlight the effectiveness of noncollinear spin-flip TDDFT in accurately locating conical intersections while mitigating spin contamination issues.

\begin{table}[htb]
    \caption{\label{tab:realci}Root Mean Square Deviations (RMSDs, in $\si{\angstrom}$) of MECI geometries with respect to MRCISD results\cite{nikiforov2014assessment} and $\expval{\hat{S}^2}$ values of $\m{S}_0$ and $\m{S}_1$ states at the MECI geometries.}
    \begin{tabular}{ccccccc}
    \hline
            & \multicolumn{3}{c}{Collinear BHHLYP\annotation{a}}   & \multicolumn{3}{c}{Noncollinear BHHLYP}  \\ \hline
    System  & RMSD  & $\expval{\hat{S}^2}_{\m{S}_0}$ & $\expval{\hat{S}^2}_{\m{S}_1}$ & RMSD  & $\expval{\hat{S}^2}_{\m{S}_0}$ & $\expval{\hat{S}^2}_{\m{S}_1}$  \\ \hline
    Ethylene& 0.077 & 0.277                    & 0.435                    & 0.018 & 0.023                    & 0.092                   \\
    PSB3    & 0.240 & 0.092                    & 0.579                    & 0.035 & 0.071                    & 0.107                   \\
    Styrene & 0.102 & 0.072                    & 0.932                    & 0.031 & 0.113                    & 0.117                   \\ \hline
    \end{tabular}

    \annotation{a}Taken from Ref.~\citenum{nikiforov2014assessment}.
\end{table}

Importantly, the noncollinear spin-flip TDDFT framework is not limited to hybrid functionals. It can be combined with various types of exchange-correlation functionals, including LDA, GGA, meta-GGA, and hybrid functionals. Appendix~\ref{appendix:BasFun} presents detailed tests on the performance of these different functionals, as well as a discussion of the basis set dependence of the MECI geometries.

To verify that noncollinear spin-flip TDDFT indeed captures the topological features of the intersection, we take the ethylene molecule as an example and plot the $\m{S}_0$ and $\m{S}_1$ PESs near the conical intersection point. As shown in Figure~\ref{fig:realci}b, the computed potential energy surfaces clearly display the characteristic double-cone topology, demonstrating that the noncollinear approach reproduces the correct intersection shape. The B3LYP functional is employed here as a representative example, though other functionals yield similar conical intersection structures.

\subsection{\label{sec:fakeci}Avoiding Spurious PES Crossings}

For pure exchange-correlation functionals, the excitation energies obtained from collinear spin-flip TDDFT are simply the energy differences between the occupied spin-up and virtual spin-down molecular orbitals. When the two singly-occupied orbitals (the O orbitals in Figure~\ref{fig:sfsch}) are (near-)degenerate, either due to symmetry-imposed degeneracy or accidental degeneracy, the resulting excitation energies can become (near-)degenerate as well, resulting in physically incorrect state degeneracies.

\begin{figure}[htb]
    \includegraphics[width=0.95\textwidth]{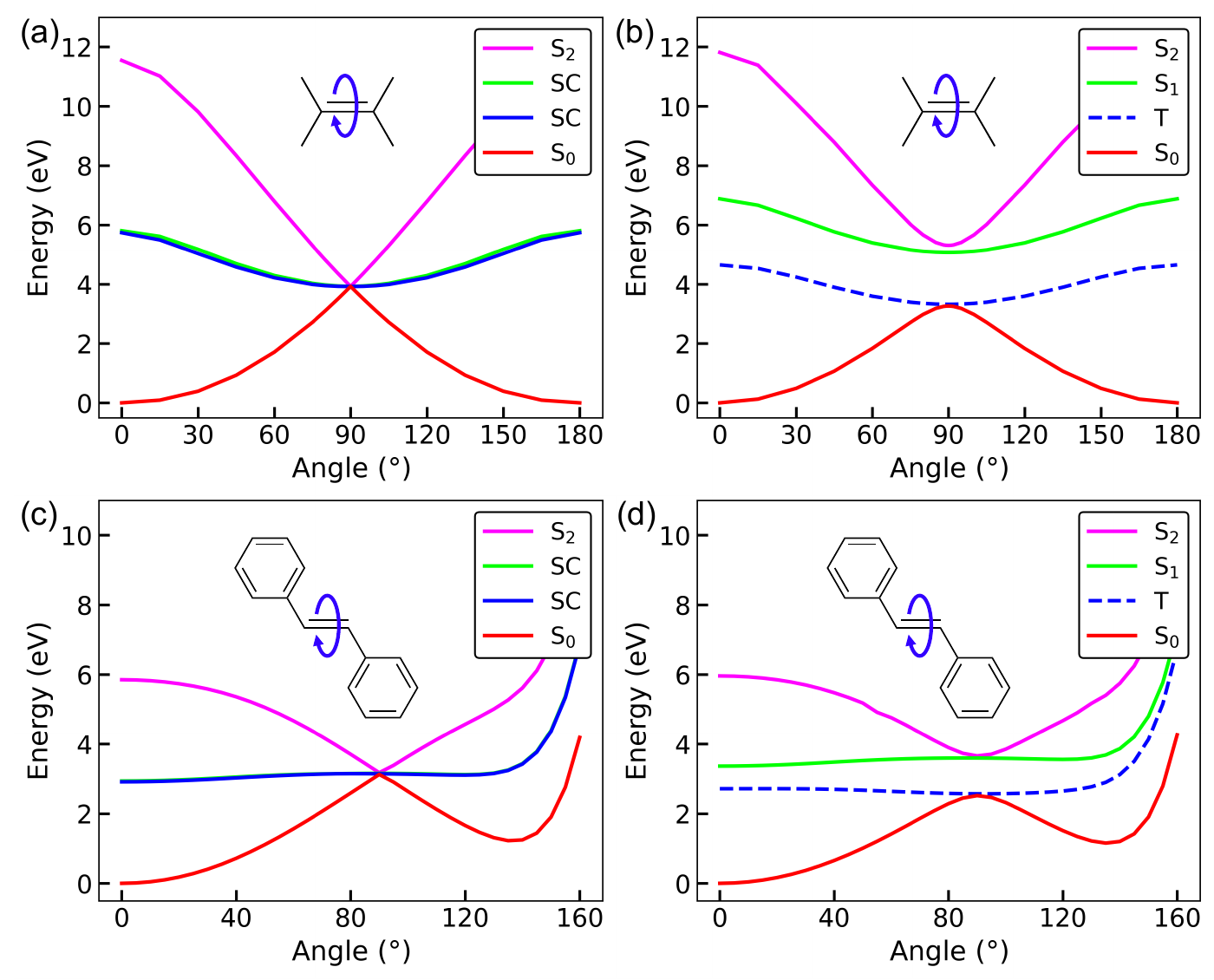}
    \caption{\label{fig:fakeci}Potential energy curves for ethylene (a, b) and stilbene (c, d) as a function of torsion angle, computed using collinear (a, c) and noncollinear (b, d) spin-flip TDDFT with BLYP functional. SC stands for strongly spin-contaminated states with $\expval{\hat{S}^2}\approx 1$. Geometries of twisted stilbene are generated using Molclus\cite{molclus}.}
\end{figure}

Figure~\ref{fig:fakeci} illustrates this problem using two examples. Figure~\ref{fig:fakeci}a shows the potential energy curves for the ground state and three low-lying excited states of ethylene as a function of the torsion angle, calculated with collinear spin-flip TDDFT (BLYP/cc-pVDZ\cite{becke1988densityfunctional,lee1988development,dunning1989gaussian}). At the $\ang{90}$ twisted geometry, the $\m{D}_{2d}$ point group imposes an exact degeneracy of frontier orbitals, which in turn enforces an artificial fourfold degeneracy predicted by collinear spin-flip TDDFT. A similar issue is observed for stilbene (Figure~\ref{fig:fakeci}c), where an accidental near-degeneracy of the frontier orbitals arises at $\ang{90}$ torsion ($\m{C}_2$ symmetry, calculated with BLYP/6-31G\cite{ditchfield1971selfconsistent,hehre1972self}). Such spurious degeneracies originate from the inherent limitation of collinear spin-flip TDDFT when using pure functionals, and are independent of the choice of basis set or functional. As pointed out by Krylov and co-workers\cite{shao2003spin}, introducing hybrid functionals into collinear spin-flip TDDFT can partially alleviate the issue to some extent by lifting finite energy gaps. However, the magnitude of these gaps can be arbitrarily tuned by adjusting the fraction of HF exchange in the functional, rendering the remedy somewhat ad hoc.

In contrast, noncollinear spin-flip TDDFT naturally lifts these degeneracies even when pure functionals are used, as shown in Figure~\ref{fig:fakeci}b and \ref{fig:fakeci}d. This demonstrates an advantage of the noncollinear approach: it inherently avoids spurious state crossings that arise from orbital degeneracies, thereby providing a more robust description of the order of electronic states. It is also worth noting that Figure~\ref{fig:fakeci}b and \ref{fig:fakeci}d closely resemble the TDDFT-1D results reported by Teh and Subotnik,\cite{teh2019simplest} where doubly excited configurations are explicitly included, highlighting the importance of double excitations. Such doubly-excited configuration $\ket{\m{DES}}$ is incorporated naturally within noncollinear spin-flip TDDFT. As a direct demonstration of this, Appendix~\ref{appendix:DES} presents several examples of excited states dominated by $\ket{\m{DES}}$.

It is worth noting that, at the $\ang{90}$ twisted geometry of ethylene and stilbene, noncollinear spin-flip TDDFT predicts near-degeneracies between $\m{S}_1$ and $\m{S}_2$, as well as between $\m{S}_0$ and $\m{T}$. Such near-degeneracies are well-acknowledged features of diradical systems\cite{pittner1999assessment,stuyver2019diradicals}.

\subsection{\label{sec:selfsplitting}Locating $\m{S}_0/\m{T}$ Crossings}

A key requirement for locating intersections between singlet and triplet states is that the spin-flip TDDFT formalism should yield a unique and consistent energy for all components of a given triplet state. In practice, however, collinear spin-flip TDDFT breaks this degeneracy, producing an artificial energy splitting of roughly 1 eV between the reference state $\ket{\m{T}(S_z=1)}$ and the spin-flip state $\ket{\m{T}(S_z=0)}$. This self-splitting problem introduces ambiguity in defining the triplet energy and consequently affects the identification of singlet–triplet crossings.

In contrast, noncollinear spin-flip TDDFT rigorously preserves the degeneracy between different $S_z$ components of the same spin multiplet, as formally proven in Ref.~\citenum{wang2025zero} and illustrated in Figure~\ref{fig:sfsch}. It is worth noting that this theorem further holds for any high-spin reference with nonzero $S_z$ (e.g., $S_z=\tfrac{1}{2}, 1, \tfrac{3}{2}, \dots$).

\begin{figure}[htb]
    \includegraphics[width=0.8\textwidth]{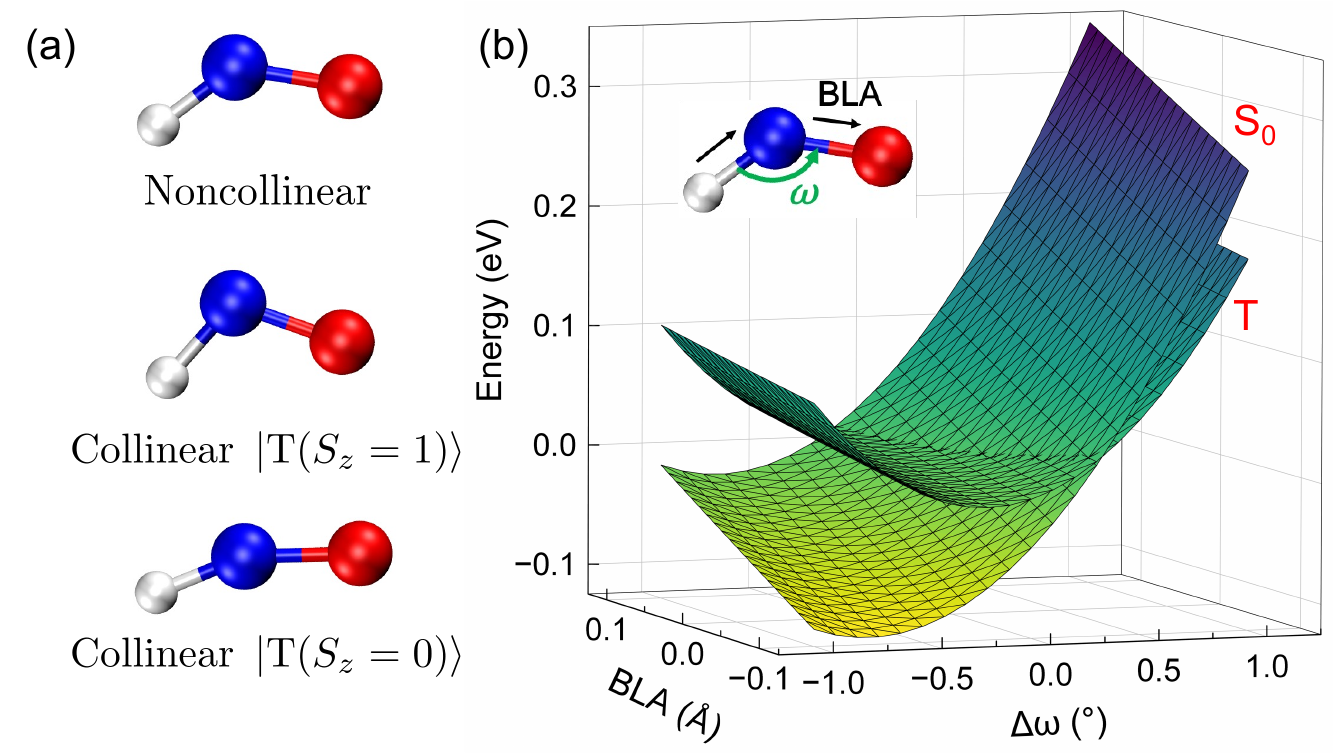}
    \caption{\label{fig:nitroxyl}(a) MECP geometries of nitroxyl calculated by different methods. (b) PESs of $\m{S}_0$ and $\m{T}$ states near the crossing point calculated by noncollinear spin-flip TDDFT. BLA stands for bond length alternation.}
\end{figure}

To demonstrate this capability, we examine the $\m{S}_0$/$\m{T}$ crossing of nitroxyl (\ch{HNO}). Figure~\ref{fig:nitroxyl}a and Table~\ref{tab:nitroxyl} summarize the MECP geometries obtained at the spin-flip TDDFT B3LYP/cc-pVTZ level. In the collinear framework, separate crossings are found depending on whether $\ket{\m{T}(S_z=1)}$ or $\ket{\m{T}(S_z=0)}$ is used as the target state, leading to markedly different geometries and energies. In contrast, noncollinear spin-flip TDDFT produces a unique MECP.

\begin{table}[htb]
    \caption{\label{tab:nitroxyl}MECP geometries of nitroxyl (\ch{HNO}) calculated by noncollinear spin-flip TDDFT, collinear spin-flip TDDFT with $\ket{\m{T}(S_z=1)}$, and collinear spin-flip TDDFT with $\ket{\m{T}(S_z=0)}$. Artificial triplet splitting energies $\Delta E=E(\ket{\m{T}(S_z=1)})-E(\ket{\m{T}(S_z=0)})$ for collinear spin-flip TDDFT are also shown.}
    \begin{tabular}{ccccc}
        \hline
        Methods       & $R$(NH) ($\si{\angstrom}$) & $R$(NO) ($\si{\angstrom}$) & $\angle$(HNO) ($\si{\degree}$) & $\Delta E$ (eV) \\ \hline
        Noncollinear  & 1.012                 & 1.224                 & 133.552              & 0.000 \\
        Collinear $\ket{\m{T}(S_z=1)}$ & 1.049                 & 1.211                 & 111.981              & 1.649 \\
        Collinear $\ket{\m{T}(S_z=0)}$ & 1.001                 & 1.216                 & 159.249              & 1.641 \\ \hline
    \end{tabular}
\end{table}

To verify that noncollinear TDDFT can provide the correct topography near the $\m{S}_0/\m{T}$ crossing point, we present the PESs near the crossing point in Figure~\ref{fig:nitroxyl}b. It can be seen that the degeneracy is only lifted along a single direction, resulting in a line-shaped crossing seam. This stands in sharp contrast to the $\m{S}_0/\m{S}_1$ conical intersection, where two effective directions lift the degeneracy.

\subsection{\label{sec:S1T}Locating $\m{S}_1/\m{T}$ Crossings}

\begin{figure}[htb]
    \includegraphics[width=0.5\textwidth]{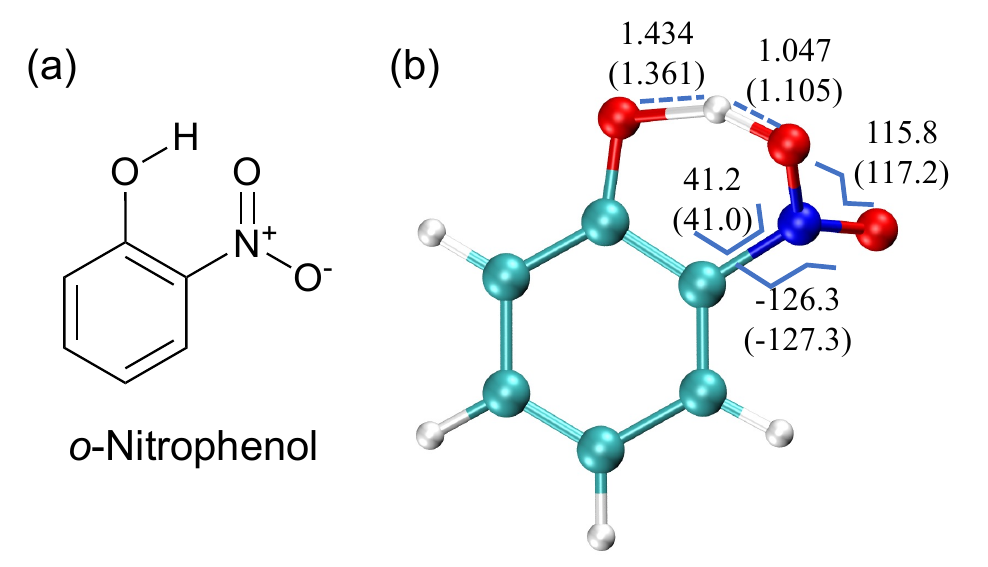}
    \caption{\label{fig:S1T}(a) Chemical structure of $o$-nitrophenol. (b) Geometry of the $\m{S}_1/\m{T}$ crossing point calculated by noncollinear spin-flip TDDFT. The numbers indicate the bond lengths (in $\si{\angstrom}$) and dihedral angles (in $\si{\degree}$), with the values in parentheses corresponding to the reference results.}
\end{figure}

Aside from the $\m{S}_0/\m{T}$ crossings, noncollinear spin-flip TDDFT can also be used to locate the $\m{S}_1/\m{T}$ crossings. As an example, $\m{S}_1/\m{T}$ crossing of $o$-nitrophenol is located at B3LYP/6-31G** level, with the structure shown in Figure~\ref{fig:S1T}. It shows that the geometry is quite similar to the reference results calculated by 6SA-CASSCF(10,10)/6-31G** in Ref.~\citenum{xu2015photoisomerization}.

It should be noted that locating the crossing point between $\m{S}_1$ and $\m{T}$ is not trivial. In particular, it requires the use of electronic structure methods that allow the energy of $\m{S}_1$ to be lower than that of $\m{T}$, which represents a violation of Hund's rule. Conventional spin-conserving TDDFT typically fails to achieve this\cite{desilva2019inverted}, whereas spin-flip TDDFT can\cite{pollice2021organic,lashkaripour2025addressing}.

\subsection{\label{sec:D0D1}$\m{D}_0/\m{D}_1$ Conical Intersections for Odd-Electron Systems}

\begin{figure}[htb]
    \includegraphics[width=0.5\textwidth]{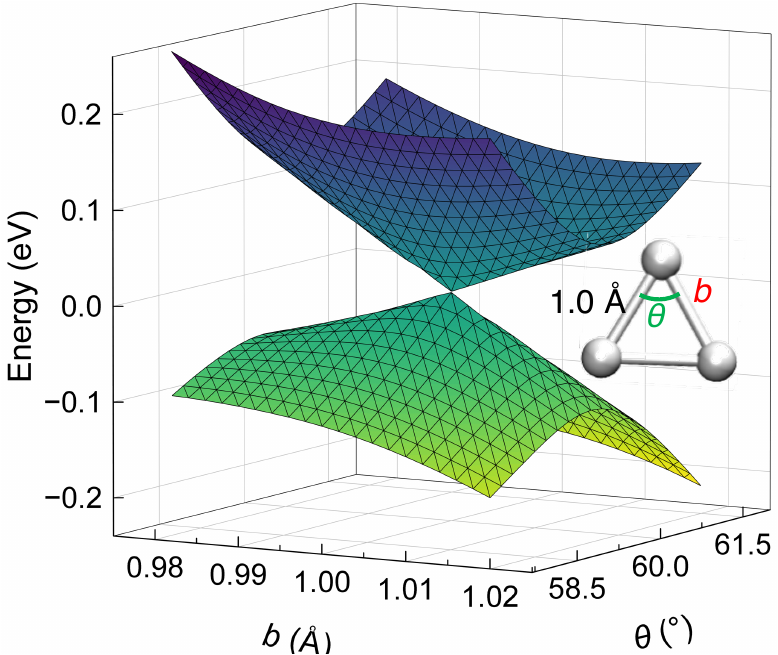}
    \caption{\label{fig:h3} PESs of $\m{D}_0$ and $\m{D}_1$ states near the conical intersection of \ch{H3} cluster.}
\end{figure}

While the preceding discussion focused on even-electron systems, the applicability of noncollinear spin-flip TDDFT extends to odd-electron molecules. To demonstrate this, we investigate the conical intersection between the two lowest doublet states ($\m{D}_0$ and $\m{D}_1$) of the \ch{H3} cluster. At the equilateral triangle geometry, the $\m{D}_{3h}$ symmetry enforces the degeneracy of these two states. The potential energy surfaces (PESs) of $\m{D}_0$ and $\m{D}_1$ are scanned using noncollinear spin-flip TDDFT. The procedure follows Ref.~\citenum{zhang2014analytic}, where one bond length is fixed at $\msi{1.0}{\angstrom}$, and the other bond length $b$ with the included angle $\theta$ are varied. As depicted in Figure~\ref{fig:h3}, the resulting surfaces clearly reveal the characteristic double-cone topology of the conical intersection.

However, it is worth noting that for systems with an odd number of electrons, conical intersections ($M-2$ dimensional intersection space) only occur in the absence of spin-orbit coupling. When spin-orbit coupling is included, each potential energy surface corresponds to a pair of time-reversal-related states sharing the same energy (unless time-reversal symmetry is broken by an external magnetic field). In such cases, the intersection space dimensionality becomes $M-5$\cite{mead1979noncrossing}.

\section{\label{sec:evaluation}Assessment of the Potential of Noncollinear Spin-Flip TDDFT for Nonadiabatic Dynamics Simulations}

Nonadiabatic dynamics simulations impose specific requirements on electronic structure methods. These requirements, which may not be exhaustive due to the authors' limited knowledge, are listed below. We evaluate the performance of noncollinear spin-flip TDDFT against each of them.

\begin{enumerate}
    \item \textbf{Description of multiple PESs with correct spin multiplicities.} Nonadiabatic dynamics relies on multiple potential energy surfaces to describe the evolution of a system, making this a prerequisite condition. Noncollinear spin-flip TDDFT naturally describes several low-lying singlet and triplet states, making it well suited for simulating nonadiabatic couplings among these states. Meanwhile, compared to collinear spin-flip TDDFT, spin contamination is significantly reduced. However, to fully eliminate spin contamination, especially for higher excited states, spin-adaptation techniques are still required\cite{li2010spinadapted,li2011spinadapted,li2011spinadapteda,zhang2015spinflip,lee2018eliminating,park2023mixedreference}.

    \item \textbf{Analytic gradients of PESs.} Efficient nonadiabatic dynamics requires analytic gradients of PESs to propagate nuclear wavefunctions. With the numerically stable spin-flip kernel (Eq.~\ref{Exx}) and its derivatives, noncollinear spin-flip TDDFT can provide analytic gradients\cite{li2025analytic}.

    \item \textbf{Correct ordering of states.} An roughly correct ordering of electronic states is important because it determines whether potential energy surface crossings exist within a given range of molecular geometries. We have shown an example in Section~\ref{sec:S1T} where, unlike spin-conserving TDDFT that always predicts the energy of $\m{S}_1$ to be higher than that of $\m{T}$, spin-flip TDDFT can meaningfully predict whether $\m{S}_1$ lies below $\m{T}$. Molecules with small or even negative $\m{S}_1/\m{T}$ gaps at their ground-state geometries, such as thermally activated delayed fluorescence (TADF) molecules, are of great significance in organic light-emitting diodes (OLEDs)\cite{desilva2019inverted,pollice2021organic,won2023inverted,perez-jimenez2025role}. The two molecules shown in Figure~\ref{fig:invest}, cyclazine and heptazine, are typical examples believed to have $\m{S}_1$ energies lower than $\m{T}$. Vertical excitation energies of two molecules are computed using both spin-conserving and spin-flip TDDFT at the geometries reported in Ref.~\citenum{loos2023heptazine}, where they were optimized at CCSD(T)/cc-pVTZ level. The results in Table~\ref{tab:invest} clearly show that noncollinear spin-flip TDDFT (except for the SVWN functional) predicts the correct ordering of $\m{S}_1$ and $\m{T}$ states, while spin-conserving TDDFT always predicts $\m{S}_1$ to lie above $\m{T}$. Finally, it is worth mentioning that degeneracy of potential energy surfaces can also be regarded as a special case of relative ordering. The degeneracy between the $\ket{\m{T}(S_z=1)}$ and $\ket{\m{T}(S_z=0)}$ is strictly preserved in noncollinear spin-flip TDDFT\cite{wang2025zero}.
    \begin{figure}[htb]
        \includegraphics[width=0.3\textwidth]{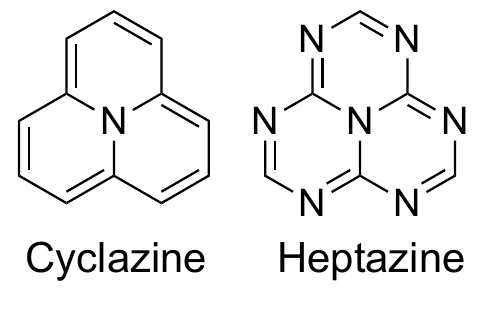}
        \caption{\label{fig:invest}Chemical structures of cyclazine and heptazine.}
    \end{figure}
    \begin{table}[htb]
        \caption{\label{tab:invest}Singlet-triplet gap $E(\m{S}_1)-E(\m{T})$ (eV) computed with different functionals using cc-pVTZ basis set.}
        \begin{tabular}{ccccc}
            \hline
            System                     & Functional & Spin-conserving & Spin-flip & Ref\annotation{a}  \\ \hline
            \multirow{5}{*}{Cyclazine} & SVWN        & 0.147           & $0.049$     & \multirow{5}{*}{$-0.131$} \\
                                    & BLYP        & 0.166           & $-0.024$    &                         \\
                                    & B3LYP       & 0.205           & $-0.240$    &                         \\
                                    & PBE         & 0.179           & $-0.019$    &                         \\
                                    & PBE0        & 0.241           & $-0.365$    &                         \\ \hline
            \multirow{5}{*}{Heptazine} & SVWN        & 0.159           & $0.034$     & \multirow{5}{*}{$-0.219$} \\
                                    & BLYP        & 0.176           & $-0.045$    &                         \\
                                    & B3LYP       & 0.220           & $-0.322$    &                         \\
                                    & PBE         & 0.187           & $-0.043$    &                         \\
                                    & PBE0        & 0.254           & $-0.481$    &                         \\ \hline
        \end{tabular}

        \annotation{a}Theoretically best estimate values from Ref.~\citenum{loos2023heptazine}.
    \end{table}

    \item \textbf{Correct topology at potential energy surface crossings.} Regions where two potential energy surfaces intersect are where nonadiabatic processes efficiently take place. Therefore, providing a topologically correct description of the potential energy surfaces near such crossing points is a critical factor for achieving reliable nonadiabatic dynamics. Noncollinear spin-flip TDDFT meets this requirement. As demonstrated in Sections~\ref{sec:ci}, it correctly describes the conical intersections between singlet states as well as the crossing seams between singlet and triplet states.

    \item \textbf{Nonadiabatic derivative coupling and spin-orbit coupling.} They are the driving forces behind nonadiabatic transitions between electronic states of the same and different spin multiplicities, respectively. Therefore, they are essential ingredients in nonadiabatic dynamics methods such as Tully's fewest switches surface hopping\cite{tully1990molecular,cui2014generalized,crespo-otero2018recent}. Similar to the treatment of analytic gradients\cite{zhang2014analytic}, noncollinear spin-flip TDDFT is capable of providing analytic nonadiabatic derivative couplings, a feature that is currently under development. However, due to slight spin contamination, nonadiabatic couplings between states of different multiplicities may be nonzero, and fully avoiding such unphysical couplings relies on appropriate spin-adaptation within the noncollinear spin-flip TDDFT framework. Moreover, one notable advantage of the noncollinear spin-flip TDDFT approach is that it offers a consistent framework for treating both derivative couplings and spin-orbit coupling effects simultaneously, which is particularly beneficial for systems where intersystem crossing and internal conversion are energetically competitive\cite{marian2012spin}.

    \item \textbf{Description of single-bond dissociation.} Nonadiabatic dynamics simulations may involve significant changes in bond lengths and angles, and even chemical reactions, in which case strong multiconfigurational character may arise. Noncollinear spin-flip TDDFT is particularly well-suited to describe such situations, as it allows the simultaneous treatment of several important configurations, including $\ket{\m{CSS}}$, $\ket{\m{OSS}}$, $\ket{\m{DES}}$. In Appendix~\ref{appendix:bonddisso}, noncollinear spin-flip TDDFT is utilized to investigate single-bond dissociation and is compared with broken-symmetry unrestricted Kohn–Sham (UKS) DFT.

    \item \textbf{Computational efficiency.} Computational cost is always a key consideration for practical dynamics simulations. As discussed in Ref.~\citenum{wang2025zero}, the computational cost of noncollinear spin-flip TDDFT and its gradients is comparable to that of DFT and spin-conserving TDDFT, significantly lower than that of traditional multireference wavefunction methods. In this work, the computational timings of noncollinear spin-flip TDDFT is further compared with collinear spin-flip TDDFT, with the results shown in Table~S5 in the Supporting Information. The noncollinear version is indeed more computationally demanding, but the difference diminishes as the system size increases.
\end{enumerate}

\section{\label{sec:conclusion}Conclusions and Outlook}

Noncollinear spin-flip TDDFT, like conventional spin-conserving TDDFT, has a rigorous theoretical foundation. With multicollinear functionals, it can correctly capture the topological features near potential energy surface crossings and provide reasonable spin multiplicities without extra approximations. Compared to traditional DFT/TDDFT methods, its advantage lies in successfully handling $\m{S}_0/\m{S}_1$ conical intersections. Compared to collinear spin-flip TDDFT, it has three main advantages: First, it significantly reduces spin contamination near conical intersections. Second, it can accurately describe singlet-triplet crossings without the self-splitting issue. Third, it does not impose specific requirements on the choice of functional, free of artificial potential energy surface crossings caused by degenerate orbitals even for pure functionals.

While the method encounters certain limitations, such as slight spin contamination in low-lying excited states and inadequate spin description for some high-lying states, its low computational cost still makes it a promising electronic structure tool for nonadiabatic dynamics simulations, especially for larger systems.

\appendix
\section*{Appendix}

\section{\label{appendix:BasFun}Basis Set and Functional Dependence of MECI Geometries}

To assess the dependence of MECI geometries on the basis set, we optimize the MECI of ethylene using noncollinear spin-flip TDDFT with BHHLYP functional and a series of basis sets\cite{ditchfield1971selfconsistent,hehre1972self,hariharan1973influence,clark1983efficient,dunning1989gaussian,kendall1992electron}. The results are summarized in Table~\ref{tab:bas}. Using the largest basis set, aug-cc-pVQZ, as the reference, RMSDs of the optimized geometries remain very small--around or below $\msi{0.01}{\angstrom}$--indicating that the choice of basis set has a negligible effect on the computed MECI geometries.

\begin{table}[htb]
    \caption{\label{tab:bas}RMSD of MECI geometries of MECI geometries of ethylene obtained by noncollinear spin-flip TDDFT (BHHLYP) using different basis sets relative to those computed using the aug-cc-pVQZ basis set.}
    \begin{tabular}{ccc}
        \hline
        Basis Set   & Number of Basis Functions & RMSD ($\si{\angstrom}$) \\ \hline
        6-31G*      & 36                        & 0.014 \\
        6-31+G**    & 56                        & 0.008 \\
        cc-pVDZ     & 48                        & 0.011 \\
        cc-pVTZ     & 116                       & 0.003 \\
        cc-pVQZ     & 230                       & 0.002 \\
        aug-cc-pVDZ & 82                        & 0.011 \\
        aug-cc-pVTZ & 184                       & 0.000 \\
        aug-cc-pVQZ & 344                       & --    \\ \hline
    \end{tabular}
\end{table}

The effect of the exchange-correlation functional is also examined by optimizing the MECIs of ethylene, PSB3, and styrene using noncollinear spin-flip TDDFT with the 6-31+G** basis set and different functionals\cite{dirac1930note,vosko1980accurate,becke1988densityfunctional,lee1988development,becke1993densityfunctional,stephens1994initio,perdew1996generalized,adamo1999reliable,yanai2004new,tao2003climbing,zhao2008m06}. The resulting geometries are compared with high-level MRCISD data\cite{nikiforov2014assessment} and the corresponding RMSDs are summarized in Table~\ref{tab:fun}. In contrast to the basis-set dependence, the choice of functional shows a more pronounced influence on the MECI geometries. Hybrid functionals generally perform better than pure functionals in reproducing the MECI geometries, with BHHLYP giving the best results among the tested functionals for the present systems. On the other hand, pure functionals tend to yield smaller spin contamination.

\begin{table}[htb]
    \caption{\label{tab:fun}RMSD (in $\si{\angstrom}$) of MECI geometries obtained by noncollinear spin-flip TDDFT with various functionals relative to MRCISD results\cite{nikiforov2014assessment} and $\expval{\hat{S}^2}$ values of $\m{S}_0$ and $\m{S}_1$ states at the MECI geometries.}
    \begin{tabular}{cccccccccc}
        \hline
        \multirow{2}{*}{Functionals} & \multicolumn{3}{c}{Ethylene}                                            & \multicolumn{3}{c}{PSB3}                                                & \multicolumn{3}{c}{Styrene}                                             \\ \cline{2-10} 
                                    & RMSD  & $\expval{\hat{S}^2}_{\m{S}_0}$ & $\expval{\hat{S}^2}_{\m{S}_1}$ & RMSD  & $\expval{\hat{S}^2}_{\m{S}_0}$ & $\expval{\hat{S}^2}_{\m{S}_1}$ & RMSD  & $\expval{\hat{S}^2}_{\m{S}_0}$ & $\expval{\hat{S}^2}_{\m{S}_1}$ \\ \hline
        SVWN                         & 0.215 & 0.008                          & 0.014                          & 0.490 & 0.009                          & 0.013                          & 0.145 & 0.011                          & 0.015                          \\
        BLYP                         & 0.201 & 0.009                          & 0.026                          & 0.375 & 0.035                          & 0.024                          & 0.126 & 0.020                          & 0.027                          \\
        B3LYP                        & 0.103 & 0.012                          & 0.050                          & 0.248 & 0.027                          & 0.050                          & 0.104 & 0.043                          & 0.046                          \\
        BHHLYP                       & 0.018 & 0.023                          & 0.092                          & 0.035 & 0.071                          & 0.107                          & 0.031 & 0.113                          & 0.117                          \\
        PBE                          & 0.188 & 0.011                          & 0.029                          & 0.335 & 0.033                          & 0.028                          & 0.116 & 0.022                          & 0.029                          \\
        PBE0                         & 0.028 & 0.017                          & 0.060                          & 0.169 & 0.037                          & 0.068                          & 0.099 & 0.057                          & 0.061                          \\
        CAM-B3LYP                    & 0.051 & 0.014                          & 0.063                          & 0.032 & 0.071                          & 0.072                          & 0.101 & 0.068                          & 0.076                          \\
        TPSS                         & 0.106 & 0.014                          & 0.043                          & 0.365 & 0.064                          & 0.041                          & 0.107 & 0.034                          & 0.040                          \\
        M06-2X                       & 0.118 & 0.016                          & 0.105                          & 0.057 & 0.100                          & 0.054                          & 0.113 & 0.037                          & 0.108                          \\ \hline
    \end{tabular}
\end{table}

\section{\label{appendix:DES}Calculations of Excited States Dominated by Doubly Excited Configuration}

\begin{figure}[htb]
    \includegraphics[width=0.5\textwidth]{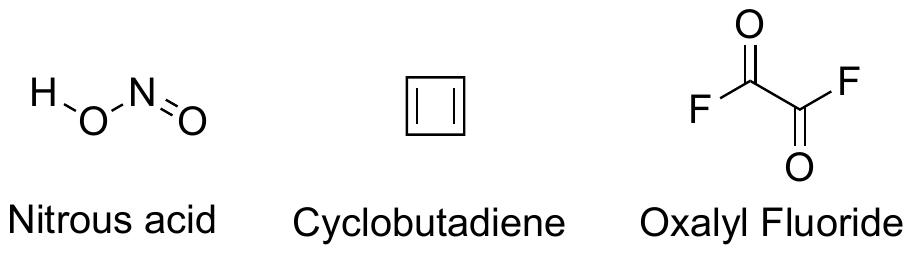}
    \caption{\label{fig:des}Chemical structures of the three test systems.}
\end{figure}

The $\ket{\m{DES}}$ configuration plays important role in shaping the topological features of $\m{S}_0/\m{S}_1$ conical intersections\cite{teh2019simplest}, avoiding artificial potential energy surface crossings\cite{teh2019simplest}, and explaining inverted singlet-triplet gaps\cite{desilva2019inverted}. Although it is missed by conventional spin-conserving TDDFT\cite{maitra2004double}, it is naturally included in noncollinear spin-flip TDDFT as a singly-excited state with respect to the reference state. A direct way to assess the impact of $\ket{\mathrm{DES}}$ is to examine electronic states dominated by it. Three molecular systems, as shown in Figure~\ref{fig:des}, are selected here for illustrative purposes. The geometry is optimized at the B3LYP/cc-pVTZ level in Gaussian 16\cite{g16} and the vertical excitation energies of states dominated by doubly excited character are computed using noncollinear spin-flip TDDFT. The results in Table~\ref{tab:des} show reasonable agreement with the reference values, indicating that noncollinear spin-flip TDDFT can provide a reliable description of states dominated by $\ket{\m{DES}}$.
\begin{table}[htb]
    \caption{\label{tab:des}Excitation energies (eV) of states dominated by doubly excited character and partition of $\ket{\m{DES}}$ configuration calculated by noncollinear spin-flip TDDFT.}
    \begin{tabular}{cccc}
        \hline
            System          & Excitation Energy (eV) & Ref (eV)            & Partition\\ \hline
            Nitrous acid    & 7.95                   & 7.97\annotation{a}  & 0.89     \\
            Cyclobutadiene  & 4.97                   & 4.04\annotation{a}  & 0.95     \\
            Oxalyl Fluoride & 9.68                   & 8.92\annotation{a}  & 0.97     \\ \hline
    \end{tabular}

    \annotation{a}Theoretically best estimate values from Ref.~\citenum{kossoski2024reference}.
\end{table}

\section{\label{appendix:bonddisso}Single Bond Dissociation}

The bond dissociation test originally reported in Ref.~\cite{wang2025zero} is revisited and slightly revised here to provide a more comprehensive assessment. The test systems include hydrogen (\ch{H2}) and hydrogen fluoride (\ch{HF}), using the same computational setup as in Ref.~\citenum{wang2025zero}. In this work, noncollinear spin-flip TDDFT results are compared with those obtained from broken-symmetry UKS DFT, with full configuration interaction (FCI) serving as the reference.

\begin{figure}[htb]
    \includegraphics[width=0.95\textwidth]{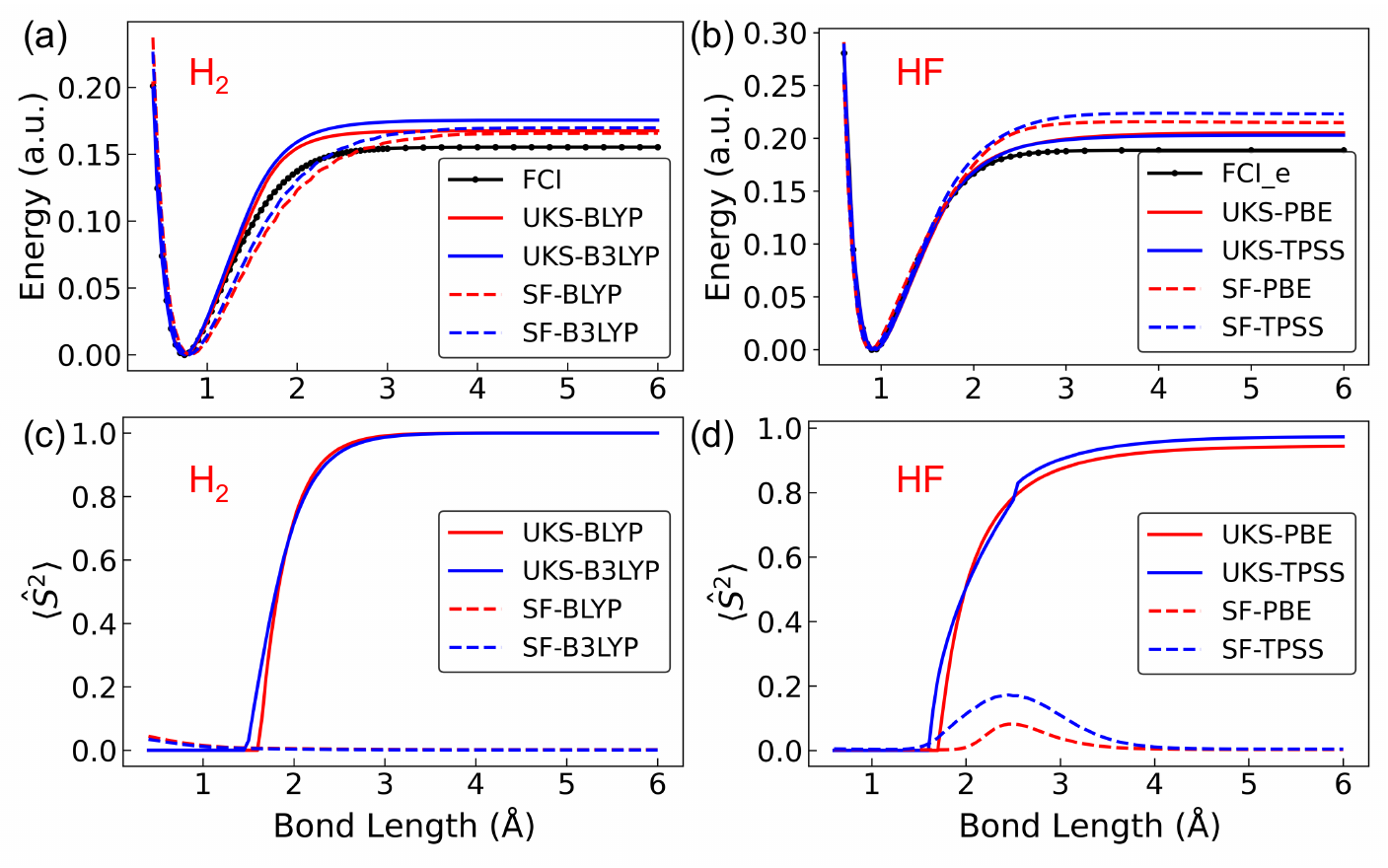}
    \caption{\label{fig:bonddisso}Potential energy curves of (a) \ch{H2} and (b) \ch{HF}, along with the corresponding $\expval{\hat{S}^2}$ values in (c) and (d), computed using noncollinear spin-flip TDDFT and UKS DFT. SF stands for spin-flip. The red and blue dashed lines in (c) are almost overlapping.}
\end{figure}

As shown in Fig.~\ref{fig:bonddisso}, both UKS and noncollinear spin-flip TDDFT yield qualitatively correct potential energy curves for the bond dissociation processes, and the UKS results even show slightly better agreement with FCI in the case of \ch{HF}. However, UKS suffers from severe spin contamination, as indicated by the large deviation of $\expval{\hat{S}^2}$ from its ideal value, whereas the noncollinear spin-flip TDDFT method exhibits only minor spin contamination. Moreover, the UKS calculations for \ch{HF} require careful SCF convergence procedures, including stability analysis, second-order SCF algorithms, and specific initial guesses, while the noncollinear spin-flip TDDFT achieves robust convergence with only a simple symmetry constraint of the $^3\Sigma$ reference state.

\begin{acknowledgement}
We thank Hao Li and Ning Zhang for valuable discussions. This work was supported by the National Key R\&D Program of China under Grant No.2025YFB3003603, the National Natural Science Foundation of China (22373004, 21927901, T2495221) and New Cornerstone Science Foundation (NCI202305).
\end{acknowledgement}

\begin{suppinfo}
The Supporting Information is available free of charge at [URL].
\begin{itemize}
\item Cartesian coordinates of MECIs and MECPs located in this work.
\item Cartesian coordinates of geometries used in Table~\ref{tab:des}.
\item Computational cost comparison between noncollinear spin-flip TDDFT and collinear spin-flip TDDFT.
\end{itemize}
\end{suppinfo}

\bibliography{ref}

\end{document}